\PassOptionsToPackage{pdfpagelabels=false}{hyperref} 
\documentclass[fleqn,usenatbib]{mnras}

\usepackage[T1]{fontenc}
\usepackage{ae,aecompl}
\usepackage{times} % comment this out if you have troubles compiling

\usepackage{ulem}
\usepackage{graphicx}	% Including figure files
\usepackage{amsmath}	% Advanced maths commands
\usepackage{amssymb}	% Extra maths symbols
\usepackage{color}
\usepackage{upgreek} % to produce \uppi (roman style for \pi which is MNRAS style)
\usepackage[utf8]{inputenc}
\usepackage{slashed}

\newcommand{\HT}{H_{\rm T}}
\newcommand{\HL}{H_{\rm L}}
\newcommand{\e}{{\rm e}}
\newcommand{\dd}{{\rm d}}
\newcommand{\be}{\begin{equation}}
\newcommand{\ee}{\end{equation}}
\newcommand{\deltaGR}{\delta_{\text{\tiny GR}}}

\title[Early radiation in N-body simulations]{The effect of early radiation in N-body simulations \\ of cosmic structure formation}

\author[J.\ Adamek et al.]{
Julian Adamek,$^{1}$\thanks{E-mail: julian.adamek@obspm.fr}
Jacob Brandbyge,$^{2,3}$
Christian Fidler,$^{4}$
Steen Hannestad,$^{2}$
\newauthor ~Cornelius Rampf\,$^{5,6}$
and Thomas Tram$^{2}$ \\
$^{1}$Laboratoire Univers et Th\'eories, Observatoire de Paris, PSL Research University, 5 place Jules Janssen, F-92195 Meudon, France\\
$^{2}$Department of Physics and Astronomy, University of Aarhus, Ny Munkegade 120, DK-8000 Aarhus C, Denmark\\
$^{3}$Centre for Star and Planet Formation, Niels Bohr Institute \& Natural History Museum of Denmark, University of Copenhagen,\\
~\,\O{}ster Volgade 5-7, DK-1350 Copenhagen, Denmark\\
$^{4}$Catholic University of Louvain -- Center for Cosmology, Particle Physics and Phenomenology (CP$^{3}$), Chemin du Cyclotron 2,\\
~\,B-1348 Louvain-la-Neuve, Belgium\\
$^{5}$Institute for Theoretical Physics (ITP), Philosophenweg 16, University of Heidelberg,  D-69120 Heidelberg, Germany \\
$^{6}$Department of Physics, Israel Institute of Technology -- Technion, Haifa 32000, Israel
}

\date{\today}

\pubyear{2017}

\begin{document}
\label{firstpage}
\pagerange{\pageref{firstpage}--\pageref{lastpage}}
\maketitle

\begin{abstract}
Newtonian N-body simulations have been employed successfully over the past decades for the simulation of the cosmological large-scale structure. Such simulations 
usually  ignore
radiation perturbations
(photons and massless neutrinos) and the impact of general relativity (GR) beyond the background expansion. This approximation can be relaxed and we discuss three different approaches that 
are accurate to leading order in GR.
For simulations that start at redshift less than about 100 we find that the presence of early radiation typically leads to percent-level effects on the numerical power
spectra at large scales.
Our numerical results agree across the three methods, and we conclude that all 
of the three methods are suitable for simulations in a standard cosmology. 
Two of the methods modify the N-body evolution directly, while the third method can be applied as a post-processing prescription.
\end{abstract}

\begin{keywords}
cosmology: theory -- large-scale structure of Universe -- dark matter
\end{keywords}

\section{Introduction}
\label{sec:intro}

Radiation plays a major role in the dynamical evolution of the early Universe. At that time, radiation and matter are tightly coupled through electroweak interactions. As the Universe expands and cools, electroweak interactions freeze out and the neutrino species decouple. The baryons maintain their tight coupling to the photons through Thomson scattering until the baryon drag epoch. From this point on, gravity is the dominating force on a range of scales, leading eventually to the gravitational collapse of matter that forms clusters and filaments -- the birth of the large-scale structure.

For the remaining
cosmological evolution, matter behaves almost as a self-gravitating system due to the difference in the dynamical time-scale of matter perturbations compared to radiation perturbations~\citep{Voruz:2013vqa}. This entails that the non-linear clustering of matter can be simulated by simply ignoring radiation perturbations, and this is what is done in Newtonian cosmological N-body simulations \citep{Teyssier:2001cp,Springel:2005mi,Hahn:2015sia}. 
On the other hand, solving the coupled Einstein--Boltzmann equations in full non-linearity on cosmologically relevant time-scales 
is currently not feasible.

However, as will be discussed in greater detail below, several recent developments make it possible to include radiation with very little computational overhead.
One example is the numerical code {\sc cosira} \citep{Brandbyge:2016raj}, which is a hybrid N-body code based on a modification of the Newtonian N-body code {\sc gadget}-2 and the linear Einstein--Boltzmann code {\sc class} \citep{Blas:2011rf}.
Here the relativistic corrections effectively appear as linear sources for the (otherwise fully non-linear) Newtonian gravity solver.

Since the effects discussed here have nearly vanishing impact on non-linear scales, instead of explicitly including these terms in a simulation,
it is also possible to account for them through post-processing of the simulation output. Such a framework has recently been developed by \citet{Fidler:2016tir}, introducing the notion
of Newtonian motion gauges.
In this framework, a modified version of {\sc class} determines the evolving space-time on which unmodified Newtonian simulations can be interpreted 
self-consistently within linear GR.
We apply this method here for the first time on actual N-body simulation data.

Other numerical techniques have been recently developed in order to carry out cosmological simulations in the context of GR. Examples include
the simulations of \citet{Bentivegna:2015flc,Giblin:2015vwq,Mertens:2015ttp} that solve the full GR evolution in a fluid approximation and focus on aspects of cosmological backreaction \citep{Buchert:2011sx,Buchert:2013qma}, a topic we shall not investigate here. To our knowledge, these codes also do not incorporate radiation and are thus not suitable for the present study.

N-body methods have many advantages if one wants to model the evolution of large-scale structure especially in the non-linear regime, in particular since they 
can handle aspects such as mergers and virialization to sufficient accuracy.
 Based on a weak-field expansion of GR, \citet{Adamek:2015eda,Adamek:2016zes} have developed the cosmological
N-body code \textit{gevolution}. Although
effects of early radiation have not been 
incorporated in the first release, we present here novel results of the most recent version 1.1 that features a linear radiation module based on the same general principle as the one employed in \textsc{cosira}.

It goes without saying that all three approaches have their benefits and drawbacks.
We will discuss and compare in detail these three approaches that go beyond the
Newtonian approximation commonly employed in a
Universe that is nowadays dominated by cold dark matter (CDM) and a cosmological constant $\Lambda$, i.e., the
$\Lambda$CDM Universe.
The paper is organized as follows: In section~\ref{sec:gauges} we will discuss the three different gauges that are at the basis of 
each method. In section~\ref{sec:codes} we describe the numerical implementation of each method and in section~\ref{sec:results} we give the results. We  conclude in section~\ref{sec:conclusions}.

\section{A tale of three gauges}
\label{sec:gauges}

In the ADM formalism \citep[after][also known as 3+1 decomposition]{Arnowitt:1962hi} a general metric is written as
\begin{equation}
\label{eq:ADM}
 \dd s^2 = -N^2 \dd\tilde{\tau}^2 + N_i N^i \dd\tilde{\tau}^2 - 2 N_i \dd\tilde{x}^i \dd\tilde{\tau} + \gamma_{ij} \dd\tilde{x}^i \dd\tilde{x}^j \,,
\end{equation}
where $N$ is the lapse function,  $N_i$ the shift vector, $\gamma_{ij}$ the metric on the three-dimensional spacelike hypersurface,
and $\tilde{\tau}$, $\tilde{x}^i$ are some arbitrary coordinates that label the foliation in the time-like direction and the points on the
hypersurfaces, respectively. We will work in the weak-field regime of GR where coordinates can be chosen such that the shift vector
is perturbatively small, and we will therefore drop the second term on the right-hand side of equation~(\ref{eq:ADM}) from now on.

We split the shift vector into a curl-free and a divergence-free component, 
\begin{equation}
 N_i = a^2 \left[\nabla_i B + B_i\right]\, , \qquad \nabla^i B_i = 0\, ,
\end{equation}
where we introduce the conformal factor $a(\tilde{\tau})$ 
which, in a Friedmann--Robertson--Walker cosmology, parametrizes the background expansion.

The three-metric $\gamma_{ij}$ can be decomposed in a similar fashion,
\begin{equation}
 \gamma_{ij} = a^2 \left[\e^{2\HL} \delta_{ij} - 2 D_{ij} \HT  + 2 \nabla_{(i} E_{j)} + h_{ij} \right] \,,
\end{equation}
where $D_{ij} = (\nabla_i \nabla_j - \delta_{ij} \nabla^2/3 )$, and 
\begin{equation}
 \nabla^i E_i = 0\, , \qquad h^i_i = 0\, , \qquad \nabla^i h_{ij} = 0 \, .
\end{equation}
The traceless part of $\gamma_{ij}$ is therefore split into a spin-0 perturbation~$\HT$, a pure spin-1 perturbation~$E_i$, and a pure spin-2 perturbation~$h_{ij}$.

Since we are working in the weak-field regime we linearize all equations in the perturbation variables $B$, $B_i$, $\HT$, $E_i$, $h_{ij}$. We
write $\exp(2 \HL)$ instead of its linearized version $(1 + 2 \HL)$ only in order to facilitate the discussion of next-to-leading order weak-field effects
later on. Similarly, we introduce a lapse perturbation $A$ by writing 
\begin{equation}
 N = a\, \e^A \,.
\end{equation} 

In GR, the freedom to choose a coordinate system implies that not all the perturbation variables introduced so far are physical. We can make a coordinate
transformation $\tau = \tilde{\tau} + T$, $x^i = \tilde{x}^i + \nabla^i L + L^i$, where $L^i$ is the divergence-free part of the spatial coordinate shift,
i.e.,
 $\nabla_i L^i = 0$. This freedom shows that two of the scalar and one of the vector perturbations (with two polarizations) defined for the metric are in
fact redundant. One way to deal with this issue is to construct a set of gauge-invariant perturbation variables, as was pioneered by \citet{Bardeen:1980kt}.
Another option is to specifically choose the coordinate system in such a way that the equations take some desired form, making them easier to solve or to interpret.
Three such choices, all relevant for cosmology and each having their own advantages and drawbacks, will be discussed in the next subsections.

\subsection{Poisson gauge}
\label{sec:Poisson}

The Poisson gauge (which in the scalar sector reduces to the so-called Newtonian or longitudinal gauge) is defined by the coordinate system where
$T$ and $L$ are chosen such that $B = \HT = 0$, and $L^i$ is chosen such that $E_i = 0$. In this case the two remaining scalar metric perturbations
$A$ and $\HL$ coincide with the two first-order gauge-invariant potentials found by \citet{Bardeen:1980kt}. We follow the notation\footnote{\label{foot:metric}In previous work concerning the N-body gauge \citep[e.g.][]{Brandbyge:2016raj,Fidler:2016tir,Fidler:2015npa} the notation of \citet{Kodama:1985bj} was used, in particular a variable
$\Phi = -\phi$. The notation of \citet{Adamek:2016zes} also features variables called $\Phi$ and $\Psi$ (stylized in capitals) but they choose a parametrization
of the metric that is due to \citet{Green:2011wc}. As will be explained later, that parametrization differs from the one employed in the present work by some coefficients of next-to-leading order terms.} of \citet{Ma:1995ey} and write $A = \psi$ and $\HL = -\phi$. The metric therefore takes the form
\begin{multline}
\label{eq:Pmetric}
 \dd s^2 = a^2 \left[-\e^{2 \psi} \dd\tau^2 - 2 B_i \dd x^i \dd \tau + ( \delta_{ij}\e^{-2 \phi}  + h_{ij} ) \dd x^i \dd x^j \right] .
\end{multline}

\citet{Adamek:2016zes} introduce a canonical momentum $q_i$ to write a geodesic equation that is valid for any value of $q_i$, including
the ultra-relativistic case $q^2 \gg m^2 a^2$. In the non-relativistic limit this becomes
\begin{equation}
\label{eq:Pgeodesic}
 \partial_\tau v_i^\mathrm{P} + \mathcal{H} v_i^\mathrm{P} = -\nabla_i \psi = -\nabla_i \phi + \nabla_i \chi\, ,
\end{equation}
where $v_i^\mathrm{P} = q_i / (m a)$ is the peculiar velocity\footnote{Note that the coordinate three-velocity is in fact
$\dd x^i / \dd \tau = \delta^{ij} (v_j^{\rm P} + B_j)$ at leading order.} in the Poisson gauge, and we have also introduced the conformal Hubble rate
$\mathcal{H} = \partial_\tau \ln a$. The lapse perturbation $\psi$ is replaced by 
\begin{equation}
\label{eq:chidef}
 \psi = \phi - \chi\, ,
\end{equation}
and one can then proceed by solving the two constraints that determine $\phi$ and $\chi$. The former is given by the Hamiltonian constraint which is to leading order
\begin{equation}
\label{eq:Pphi}
 \nabla^2 \phi - 3 \mathcal{H} \partial_\tau \phi - 3 \mathcal{H}^2 \left(\phi - \chi\right) = 4 \uppi G a^2 \sum_\alpha \bar{\rho}_\alpha \delta_\alpha^\mathrm{P}\,.
\end{equation}
Here, $\delta_\alpha^\mathrm{P} \equiv (\rho_\alpha^{\rm P} - \bar \rho_\alpha)/\bar \rho_\alpha$ are the density perturbations in the Poisson gauge
superposed on the background density $\bar \rho_\alpha$, where
$\alpha$ labels the various species, i.e., baryons, dark matter, neutrinos and photons. As opposed to the N-body gauge, discussed below, one has to keep in mind that the spatial volume is perturbed
by the presence of $\HL = -\phi$ in this gauge, which has an effect on how the physical density should be 
interpreted in 
numerical simulations.
As a related note,
mass conservation for non-relativistic species 
reads
\begin{equation}
  \partial_\tau \delta^\mathrm{P} + \nabla^i v_i^\mathrm{P} = 3 \partial_\tau \phi 
\end{equation}
at leading order, including a source term due to the local volume deformation. 

The second constraint is, to the leading order, given by
\begin{equation}
\label{eq:chi}
 \nabla^2 \nabla^2 \chi = 12 \uppi G a^2  D^{ij} \sum_\alpha \delta_{jk}\left[T_i^k\right]_\alpha\, ,
\end{equation}
where the right-hand side is the longitudinal projection of the (traceless) stress-energy commonly known as the scalar anisotropic stress. This quantity is
gauge-invariant at first order. We use the notation $\left[T_i^k\right]_\alpha$ to denote the contribution of species $\alpha$ to a tensorial quantity.

In standard cosmology the scalar anisotropic stress appears at first order only for relativistic species (photons and neutrinos) and decays rapidly inside
the horizon. On small scales, and in particular after the end of radiation domination, second-order effects from non-relativistic species and geometry
can become larger than this first-order contribution. \citet{Adamek:2016zes} therefore compute the right-hand side of equation (\ref{eq:chi}) non-perturbatively
from the N-body ensemble, and they take into account second-order geometric corrections as well, i.e.\ terms that arise from the Einstein tensor beyond leading order. 

The form of the second-order geometric terms depends on how one parametrizes the metric at second order. While \citet{Adamek:2016zes} follow the parametrization
of \citet{Green:2011wc} in writing, for instance, the lapse as $N^2 = a^2 (1 + 2 \Psi)$ even at second order, we choose to employ the more common parametrization
$N^2 = a^2 \exp(2\psi) = a^2 (1 + 2 \psi + 2 \psi^2 + \ldots)$. It should be clear that the two parameters $\psi$ and $\Psi$ can always be directly related in a
weakly perturbed geometry. Including the second-order geometric corrections, equation (\ref{eq:chi}) becomes
\begin{equation}
\label{eq:chi2nd}
 \nabla^2 \nabla^2 \chi = 3 D^{ij} \biggl(\nabla_i \phi \nabla_j \phi 
  + 4 \uppi G a^2 \sum_\alpha \delta_{jk} \left[T_i^k\right]_\alpha\biggr)\, ,
\end{equation}
up to quadratic terms involving $\chi$, $B_i$ or $h_{ij}$. The reason why we choose to neglect those terms is the fact that at leading order $\phi$ and $\psi$
are typically much larger than the other metric perturbations in Poisson gauge, and $\chi$ is very small inside the horizon. We therefore expect the
first term on the right-hand side 
in equation~(\ref{eq:chi2nd}) to give the dominant geometric contribution. A perturbative calculation
of the second-order contributions to $\chi$ is presented in appendix \ref{app:chi}.

By augmenting also the other equations we can establish accuracy at next-to-leading weak-field order on small scales for the entire scheme. The Hamiltonian
constraint becomes
\begin{multline}
\label{eq:Pphi2nd}
 \left(1 + 2 \phi\right) \nabla^2 \phi - 3 \mathcal{H} \partial_\tau \phi - 3 \mathcal{H}^2 \left(\phi - \chi\right) - \frac{1}{2} \nabla_i \phi \nabla^i \phi \\
 = 4 \uppi G a^2 \sum_\alpha \bar{\rho}_\alpha \delta_\alpha^\mathrm{P}\, ,
\end{multline}
and the geodesic equation acquires the frame-dragging term,
\begin{equation}
\label{eq:Pgeodesic2nd}
 \partial_\tau v_i^\mathrm{P} + \mathcal{H} v_i^\mathrm{P} = -\nabla_i \phi + \nabla_i \chi - v_j^\mathrm{P} \delta^{jk} \nabla_i B_k\, .
\end{equation}
The latter is still written in a low-velocity expansion in order to highlight the terms relevant for non-relativistic particles.

The frame-dragging potential $B_i$ can be obtained by extracting the divergence-free part of the momentum constraint,
\begin{equation}
\label{eq:B}
 \frac{1}{4} \nabla^2 \nabla^2 B_i = 4 \uppi G a^2 \delta_{ij} \left(\nabla^j \nabla^k - \delta^{jk} \nabla^2\right) \sum_\alpha \left[T_k^0\right]_\alpha\, .
\end{equation}
In linear perturbation theory the right-hand side is decaying and usually assumed to be negligible at the end of radiation domination. At second order the formation of
cosmic large-scale structure induces a growing frame-dragging potential \citep[e.g.][]{Lu:2008ju}. In an N-body scheme the right-hand side of the momentum constraint
can be computed non-perturbatively, a method employed for the first time by \citet{BrunI:2013mua}. In the standard model however, the effect of frame dragging on the
trajectories of non-relativistic particles remains minuscule: \citet{Adamek:2015eda} found the
change in velocity accumulated due to this effect over the
lifetime of the Universe to be typically around 10\,m\,s$^{-1}$ at megaparsec scales, five orders of magnitude\footnote{It is interesting to note that the
frame-dragging potential $B_i$ itself is only about two orders of magnitude smaller than the gravitational potentials $\psi$ and $\phi$. The additional three
orders of magnitude suppression is due to the velocity component that contracts one of the indices in the frame-dragging term of equation~(\ref{eq:Pgeodesic2nd}).
The suppression is therefore much less severe for relativistic particles.} smaller than the typical velocities of 1000\,km\,s$^{-1}$.

The spin-2 perturbation $h_{ij}$ obeys a damped wave equation with a source that is given by the spin-2 part of the anisotropic stress
\citep[e.g.][]{Adamek:2016zes}. However, the scattering of non-relativistic particles with gravitational waves is so weak that we can certainly neglect it
in N-body simulations.

\subsection{N-body gauge}
\label{sec:N-body}
Seeking a coordinate system where the equations of motion for non-relativistic matter in linearized GR are closely related to their Newtonian counterparts, \citet{Fidler:2015npa} discovered the N-body (Nb) gauge. So far it has only been discussed in the scalar sector and to first order, with coordinates chosen such that \mbox{$\HL = 0$,
thereby eliminating the spatial volume perturbation.}
In order to cast the equations into the desired form one furthermore has to set \mbox{$B = v$,} where $v$ is the total velocity potential of the
combined fluid ($v^i= \nabla^iv$). In the
scalar sector the line element reads 
\begin{multline}
\label{eq:Nmetric}
\dd s^2 =  a^2 \Big[ - {\rm e}^{2 \xi} \dd \tau^2  -2  \nabla_i B \,\dd x^i \dd \tau  \\               + \left( \delta_{ij} - 2 D_{ij} H_{\rm T}^{\rm Nb} \right) \dd x^i \dd x^j \Big] \,.
\end{multline}
Here $A=\xi$ is a linear perturbation sourced by radiation pressure and anisotropic stress;  $\xi$ thus grows in the radiation dominated era, whereas it decays in the matter and $\Lambda$-dominated eras. Furthermore, $\xi$ is in any comoving-orthogonal gauge
given by
\be
  ( \rho +  p) \xi =  - \left( 12 \uppi G a^2 \right)^{-1} \nabla^2 \chi - \delta\! p \,, 
\ee
where $\rho$, $p$ and $\delta\! p$ are respectively the density,
the pressure, and
the pressure perturbations from all fluids.

We could naturally extend this gauge to the vector sector by setting again $E_i = 0$. The remaining vector perturbation $B_i$ is then determined by the
divergence-free part of the momentum constraint just as in Poisson gauge, but it does not play any role in the dynamics when one works in the Newtonian limit.
Neither does the tensor perturbation~$h_{ij}$.

At leading order in the Nb gauge, mass conservation of the non-relativistic species is identical to the one in Newtonian gravity, i.e.,  
\be
 \partial_\tau \delta^{\rm Nb} + \nabla^i v_i^{\rm Nb} = 0 \,,
\ee
and the geodesic equation
reads
\begin{equation}
\label{eq:Ngeodesic}
\partial_\tau v_i^\mathrm{Nb} + \mathcal{H} v_i^\mathrm{Nb} = -\nabla_i \phi + \nabla_i \gamma^{\rm Nb} \,,
\end{equation}
where $\phi$ is the same gauge-invariant potential as introduced above.
It satisfies the Poisson equation
\begin{equation}
\label{eq:Nphi}
 \nabla^2 \phi = 4 \uppi G a^2 \sum_\alpha \bar{\rho}_\alpha \delta_\alpha^\mathrm{Nb}\, ,
\end{equation}
where $\delta_\alpha^\mathrm{Nb}$ are density perturbations in the Nb gauge. One important aspect of this gauge is the fact that the volume perturbation
$\HL$ is absent, and therefore, the $\delta^\mathrm{Nb}$ for non-relativistic particles is simply the counting density, i.e., the density computed
in a Newtonian N-body simulation. Furthermore, instead of the somewhat cumbersome equation~(\ref{eq:Pphi}) the potential $\phi$ is solved from the much simpler
equation~(\ref{eq:Nphi}) that directly resembles its Newtonian counterpart.

The relativistic correction $\gamma^{\rm Nb}$ to the geodesic equation is given by
\begin{equation}
\label{eq:gamma}
 \gamma^{\rm Nb} = \partial_\tau^2 \HT^{\rm Nb} + \mathcal{H} \partial_\tau \HT^{\rm Nb} + \chi\, , 
\end{equation}
where $\chi$ is the first-order gauge-invariant quantity computed in equation (\ref{eq:chi}).

In the late Universe radiation becomes less and less important. 
The contribution of radiation to the Poisson equation~(\ref{eq:Nphi}) becomes eventually negligible, and it can be shown that, to the leading weak-field order, $\gamma^{\rm Nb}$ tends to zero in that limit.
 Under these conditions the dynamical equations take
the Newtonian form, justifying the definition and the naming of the Nb gauge. However, at early times when simulations are usually initialized ($z\geq 50$), radiation remnants contaminate the evolution equations of non-relativistic particles, through the $\gamma^{\rm Nb}$ term in the geodesic equation~(\ref{eq:Ngeodesic}) and the non-matter source terms in the Poisson equation~(\ref{eq:Nphi}).
By keeping track of these terms, one can thus compute the relativistic correction
to the
Newtonian trajectories in order to recover the relativistic evolution.

\subsection{Newtonian motion gauges}
\label{sec:NMgauge}
While the Poisson and Nb gauges employ a simple gauge fixing
 -- either by directly relating the metric potentials or setting them to zero -- the Newtonian motion (Nm) gauge employs a more complex gauge definition that is equivalent to a differential equation for the metric potential $H_{\rm T}$. However, this allows us to define a gauge in which the relativistic trajectories of cold matter coincide directly with the Newtonian trajectories.
Using such a gauge makes it possible to incorporate relativistic corrections without modifying the Newtonian simulation \citep{Fidler:2016tir}.

In fact, the Nm gauges describe an entire class of gauges, 
and in this paper we choose the specific Nm gauge that employs the Nb gauge time coordinate, i.e. it is comoving-orthogonal with $B=v$ and $A= \xi$. The resulting line element for scalar perturbations is
\begin{multline}
\label{eq:NMmetric}
\dd s^2 =  a^2 \Big[ - {\rm e}^{2 \xi} \dd \tau^2  -2  \nabla_i B\, \dd x^i \dd \tau  \\               + \left( {\rm e}^{2 H_{\rm L}^{\rm Nm}} \delta_{ij} - 2 D_{ij} H_{\rm T}^{\rm Nm} \right) \dd x^i \dd x^j \Big] \,.
\end{multline}
Mass conservation is given by
\begin{equation}
\label{eq:contiNM}
   \partial_\tau \delta^{\rm Nm} + \nabla^i v_i^{\rm Nm} = -3 \partial_\tau H_{\rm L}^{\rm Nm} \,,
\end{equation}
whereas the force term in the geodesic equation is by definition identical to the Newtonian one,
\begin{equation}
 \label{eq:NMgeodesic}
\partial_\tau v_i^\mathrm{Nm} + \mathcal{H} v_i^\mathrm{Nm} = -\nabla_i \Phi^{\rm N} \,,
\end{equation}
where we have made use of the spatial gauge condition of the Nm gauge 
\begin{equation}
 \label{eq:NMspatial}
  - \Phi^{\rm N} = - \phi + \gamma^{\rm Nm} \,, 
\end{equation}
with $\gamma^{\rm Nm} =\partial_\tau^2 \HT^{\rm Nm} + \mathcal{H} \partial_\tau \HT^{\rm Nm} + \chi$ and the Newtonian potential based on the counting density $\delta^{\rm N}$
\begin{equation}
  \nabla^2 \Phi^{\rm N} = 4 \uppi G a^2 \bar \rho \delta^{\rm N} \,.
\end{equation}
Explicitly, these equations are valid for a multifluid universe, although
 the Newtonian potential is only sourced by the non-relativistic species.
Since the Nm gauge has a non-vanishing volume deformation the relativistic density is constructed from the counting density and the volume deformation
\begin{equation}
\label{eq:bookDelta}
 \delta^{\rm N} = \delta^{\rm Nm} + 3 H_{\rm L}^{\rm Nm} \,.
\end{equation}
Inserting this relation
into equation~(\ref{eq:contiNM}), we find that the variables $\Phi^{\rm N}$, $v^\mathrm{Nm}$ and $\delta^{\rm N}$ follow entirely Newtonian equations of motion and can be identified with the perturbations evolved in a Newtonian simulation. This implies that an unmodified Newtonian N-body simulation 
is in fact computing the relativistic evolution of the particles in the 
Nm gauge. We are thus able to obtain a relativistic interpretation of such an unmodified simulation by embedding its output in the nontrivial space-time of the Nm gauge. Note that the relativistic density $\delta^\mathrm{Nm}$ is not evolved by Newtonian equations and is affected by the non-trivial local volume deformation. After solving for the metric perturbations, however, we recover
the relativistic solution for the density $\delta^{\rm Nm}$ from the simulation density $\delta^{\rm N}$ by employing equation~(\ref{eq:bookDelta}).

The spatial gauge condition of the Nm gauge, equation~(\ref{eq:NMspatial}), is in fact a second-order time-differential equation, i.e.,
\be
  \partial_\tau^2 H_{\rm T}^{\rm Nm} + {\cal H} \partial_\tau H_{\rm T}^{\rm Nm} = \phi  - \chi - \Phi^{\rm N} \,.
\ee
To solve the differential equation one requires boundary conditions. 
In this work we choose the method explained in~\citet{Fidler:2016tir} for fixing these boundary conditions, corresponding to a
metric associated with a Newtonian simulation initialized in the Nb gauge. 

The metric perturbations in the Nm gauge, cf.\ equation~(\ref{eq:NMmetric}),
can be solved by modifying conventional Einstein--Boltzmann solvers such as {\sc class} or {\sc camb} \citep{Lewis:1999bs}.
Details on the explicit numerical implementation are given in the following section.

\section{Numerical implementation}
\label{sec:codes}

In the following we discuss two numerical codes, {\sc cosira} in section~\ref{sec:COSIRA} and {\it gevolution} in section~\ref{sec:gevolution},
that aim, amongst other things, to incorporate the gravitational coupling of radiation perturbations to non-relativistic matter. Then in section~\ref{sec:NewtonianMotion gauges} we summarize the steps needed to apply the Nm gauge framework 
to a Newtonian N-body simulation.

\subsection{ COSIRA (N-body gauge)}
\label{sec:COSIRA}

The code {\sc cosira} (COsmological SImulations with RAdiation) has been recently introduced as the first hybrid code incorporating 
relativistic corrections to matter trajectories in cosmological simulations \citep{Brandbyge:2016raj}.
{\sc cosira} is essentially a modified version of the Newtonian N-body code {\sc gadget-2}, a code which is originally designed to only evolve matter particles. 
In {\sc cosira}, the N-body code is interfaced with a modified version of the Einstein--Boltzmann solver {\sc class}. In this way the gravitational effect of evolving radiation perturbations on the matter N-body particles can be taken into account.

Computations are performed in the Nb gauge, and in that gauge the impact of GR can be combined into an effective GR density perturbation  $\deltaGR$ 
defined in the next paragraph. This GR density  
is added to the CDM density and passed to the Poisson solver of \mbox{{\sc gadget-2}}.
This modifies the long-range particle-mesh part of the force whereas the short-range tree part of the force is left unchanged. 

In detail, we can write for the combined force potential in the geodesic equation~(\ref{eq:Ngeodesic}), that
\begin{subequations}
\begin{equation}
 - \phi + \gamma =  - \phi_{\rm sim} - \phi_{\text{\tiny GR}} \,,
\end{equation}
with
\begin{align}
  &\nabla^2 \phi_{\rm sim}\!=   4\uppi G  a^2  \bar \rho_{\rm cdm} \delta_{\rm sim} \,, \\
  &\nabla^2 \phi_{\text{\tiny GR}} \hspace{0.0296cm}  \!=  4\uppi G  a^2  \bar \rho_{\rm cdm} \deltaGR \,, \label{def:deltaGR}
\end{align}
\end{subequations}
where \mbox{$\bar \rho_{\rm cdm} \deltaGR \equiv \sum_{\alpha\neq \rm cdm} \bar \rho_\alpha \delta_\alpha - (4\uppi Ga^2)^{-1}\nabla^2 \gamma$}, 
and $\delta_{\rm sim}$ is the non-linear CDM density contrast obtained from 
the N-body simulation. We note that in this way {\sc cosira} includes the linear effect from GR and radiation that is passed from the Einstein--Boltzmann code to the N-body code, but there is no feedback in the opposite direction. Such a feedback would occur as a non-linear
correction that goes beyond the linear scheme employed in the Einstein--Boltzmann code.

Once $\deltaGR$ has been incorporated in {\sc gadget}-2,
the code evolves matter in full non-linearity
whilst being in accordance with leading order GR. This also implies that the N-body output of {\sc cosira} should be interpreted on the Nb gauge space-time.

As mentioned above, the Nb gauge  
is so far only defined to the leading order in GR.
 As a consequence, {\sc cosira} neglects some second-order GR corrections that {\it gevolution} does take into account. This applies for instance to the second-order anisotropic stress that contributes to $\chi$ in the non-linear regime as we shall discuss in section~\ref{sec:aniso}.

\subsection{gevolution}
\label{sec:gevolution}

\citet{Adamek:2015eda,Adamek:2016zes} introduced \textit{gevolution}, the first cosmological N-body code that is based entirely on a weak-field expansion of GR.
The two main differences to the traditional Newtonian method, which in some sense is also a weak-field limit of GR, are the following:

\begin{enumerate}
 \item In Poisson gauge, the code explicitly computes all six metric perturbations, i.e.\ the two potentials $\phi$ and $\psi$, the two spin-1 modes
 of $B_i$, and the two spin-2 modes of $h_{ij}$. No assumption about the stress-energy is made except for the requirement that the gravitational
 fields have to remain small on the scales resolved by the simulation. This provides great flexibility as one can consistently include many types of relativistic
 sources for which a Newtonian treatment would be inappropriate. The example relevant for this work is, of course, radiation.
 \item The geodesic equation is solved using a relativistic canonical momentum such that arbitrary momenta are allowed (in particular the ultra-relativistic
 limit $q^2 \gg m^2 a^2$).
\end{enumerate}

One limitation of the code is the fact that it works at fixed spatial resolution, mainly because the metric perturbations are solved
using spectral analysis. However, a fixed lattice has advantages for parallelization, and \textit{gevolution} is therefore typically an order of magnitude
faster than an adaptive code for the same problem size. Furthermore, it is straightforward to add a linear source term for which a transfer function can be
computed using an Einstein--Boltzmann solver.

The most recent version 1.1 of \textit{gevolution}, presented here for the first time, can be linked directly with the Einstein--Boltzmann code {\sc class} such
that the relevant transfer functions can be computed on the fly. At each time step a realization of the linear density field of radiation (or any other linear
source such as e.g.\ light neutrinos) is prepared and added to the N-body (matter) source to obtain the right-hand side of equation (\ref{eq:Pphi2nd}). The
same is done for the anisotropic stress that sources equation (\ref{eq:chi2nd}). Linear vector modes are not generated in standard cosmology and therefore
frame dragging is only caused by non-linear
 matter according to equation (\ref{eq:B}).

It is also possible to run a Newtonian simulation with \textit{gevolution}. In this mode the evolution is performed in Nb gauge. Similar to how it is done in
the Poisson-gauge case, but in contrast to {\sc cosira}, the plain linear radiation source (in Nb gauge) is used in the Poisson equation (\ref{eq:Nphi})
such that the potential $\phi$ is computed explicitly. A realization of $\gamma$ is then prepared separately for solving the geodesic equation (\ref{eq:Ngeodesic}).
We remind the reader that {\sc cosira} instead solves directly for the combination $\gamma - \phi$ using a modified Poisson equation.

\subsection{Newtonian motion gauge and Newtonian simulations}
\label{sec:NewtonianMotion gauges}

Instead of employing a relativistic N-body simulation, the idea of the Nm gauge framework is to make use of a relativistic coordinate
system that is compatible with ordinary Newtonian N-body simulations. In this picture a Newtonian code is in fact computing the
relativistic evolution of the particles in the corresponding Nm gauge, and we are able to obtain a relativistic interpretation of this simulation by
embedding its output in the non-trivial space-time of the Newtonian motion gauge.

The method is compatible with any Newtonian 
simulation and does not require modifications of the corresponding code. In order to obtain results that can be compared with the direct implementation in the Nb~gauge 
(or Poisson gauge) presented in the previous sections, however, we need to perform a gauge transformation to the respective reference gauge. 
 This means, at the level of the final N-body output, that we displace the positions of the simulated particles to obtain the relativistic output in the 
Nb~gauge (or Poisson gauge).

Since the Nm and Nb gauge employ the same time foliation, the required gauge transformation is purely spatial, 
\begin{subequations}
\begin{align} 
  \tau_\mathrm{Nb} &= \tau_\mathrm{Nm} \,, \\
  x_\mathrm{Nb}^i &= x_\mathrm{Nm}^i + \nabla^i L_{\mathrm{Nm}\rightarrow\mathrm{Nb}}  \,, \label{spatialTrafo}
\end{align}
\end{subequations}
where $L_{\mathrm{Nm}\rightarrow\mathrm{Nb}} = 3 \HL^{\rm Nm}$.
Recall that we have chosen to initialize our simulations in Nm coordinates that agree initially with the Nb coordinates. The Nb gauge enforces $\HL^{\rm Nb}=0$, but
the presence of radiation during the simulation run generates non-zero values of $\HL$ in the Nm gauge. 
Thus, the spatial gauge transformation~(\ref{spatialTrafo}) to the Nb gauge resets $\HL^{\rm Nm} \neq 0$ to $\HL^{\rm Nb}=0$, which can be verified by the standard methods of cosmological perturbation theory \citep[see, e.g.,][]{Villa:2015ppa}.
Further details on this gauge transformation and its computation in \textsc{class} can be found in \citet{Fidler:2016tir}.

We compute the potential of the displacement field $L_{\mathrm{Nm}\rightarrow\mathrm{Nb}}$ corresponding to this gauge transformation in our modified version of \textsc{class}. Since {\it gevolution} can be run in Newtonian mode, we employ the output of {\it gevolution} in this mode and pass the information of the gauge transformation from {\sc class} to {\it gevolution}. The gauge transformation is then implemented on the final output providing the particle positions in the Nb gauge.

The advantage of this method is that the step-size in the N-body simulation is unaffected by the small time-scale of the radiation perturbations which is taken into account consistently in the Einstein--Boltzmann solver.
The disadvantage is that the non-linear growth of fluctuations induced from the radiation perturbations themselves are also included at the level of the space-time which is computed only to linear order. However, as discussed in previous works~\citep{Brandbyge:2016raj,Fidler:2016tir} and shown by the results presented in this paper, these corrections are negligible at least in a standard model cosmology.

\section{Results}
\label{sec:results}

\citet{Brandbyge:2016raj} carried out a set of N-body simulations with \textsc{cosira}. 
These use a comoving
box of size $16384$ Mpc$/h$ that contain $1024^3$ particles. The cosmology is chosen as follows: $\Omega_\mathrm{m} = 0.3133$,
$\Omega_\mathrm{b} = 0.0490$, $h = 0.6731$, $n_s = 0.9655$, ($k_\text{pivot}=0.05$\,Mpc$^{-1}$), $A_s = 2.215 \cdot 10^{-9}$, $T_\mathrm{CMB} = 2.7255$\,K,
$N_\mathrm{eff} = 3.046$ (three massless neutrinos). To compare with these existing runs, we perform a new set of N-body simulations with \textit{gevolution}
using the same cosmological parameters and the same initial redshift $z_\mathrm{ini} = 99$. Here we choose a regular lattice of $2048^3$ grid points and the
same number of particles in order to compensate for the lack of adaptive force resolution -- \textit{gevolution} employs a particle-mesh scheme on a
fixed regular mesh while \textsc{cosira} inherits the TreePM algorithm from \textsc{Gadget-2} that resolves the (Newtonian) force at sub-grid scales.
We have also run simulations with 
$512^3$, $1024^3$, and even $4096^3$ grid points and particles, 
to investigate thoroughly the convergence of our results.

\subsection{Overview of simulations}
We explore various choices of handling early radiation effects, summarized in table \ref{tab:sims}:

\begin{enumerate}
 \item[(I)] Using the `Newtonian' mode of \textit{gevolution}, we include the effect of radiation perturbations in the same manner as
 it is done in {\sc cosira}. This yields a relativistic simulation output in the Nb gauge.
 \item[(II)] Using the `relativistic' mode of \textit{gevolution} that employs Poisson gauge, we include the effects of radiation perturbations in an analogous manner,
 but adapted to Poisson gauge as explained in section \ref{sec:Poisson}. However, in order to compare the final result with the other runs, we have to convert
 it from Poisson gauge to Nb gauge. This is done by actively displacing the particles at final time according to a linear displacement field $L_{\mathrm{P}\rightarrow\mathrm{Nb}}$, the spatial gauge generator that connects
 the two gauges.
 \item[(III)] We run \textit{gevolution} in `Newtonian' mode, but instead of adding radiation perturbations in the simulation, we retroactively include radiation when interpreting the output in the Nm gauge. 
In order to compare the final result we convert it from Nm gauge to the Nb gauge by a gauge transformation, connected by the displacement field $L_{\mathrm{Nm}\rightarrow\mathrm{Nb}}$.
 \item[(IV)] We run {\it gevolution} in `relativistic' mode employing the Poisson gauge, but we completely neglect the presence of radiation perturbations. The final result is brought back to
 Nb gauge in the same way as for case (II). Note that this run does include other relativistic contributions in the Poisson gauge according to the weak-field expansion employed in {\it gevolution}. 
 \item[(V)] We run {\it gevolution} in `Newtonian' mode
neglecting the presence of radiation perturbations. As the Nb gauge equations in this limit are entirely Newtonian, this corresponds to a relativistic simulation in the Nb gauge when neglecting radiation perturbations. 
 \item [(VI)] We finally run another simulation in `Newtonian' mode with {\it gevolution} where radiation perturbations are ignored in the evolution, however, starting on initial data that
 was designed to obtain the correct power spectrum at redshift $z = 0$.
 For that purpose,
 the linear transfer functions are scaled back from $z = 0$ to the initial
 redshift using the linear (growing-mode) solution of matter that would be obtained if radiation was only present in the background. In practice this is one
 of the most common ways to deal with the effect of early radiation, and in the following we call this method ``backscaling.''
 See~\citet{Fidler:2017ebh} for a thorough theoretical discussion on this method in terms of the Nb and Nm gauges.
\end{enumerate}

\begin{table}
\caption{\label{tab:sims} Summary of simulations performed in order to study different possibilities of handling early radiation.}
\begin{tabular}{@{}ccc}
 \hline
 simulation & gauge & radiation treatment \\
 \hline
 I & N-body & direct simulation\\
 II & Poisson & direct simulation\\
 III & Newtonian motion & post-processing\\
 IV & Poisson & none\\
 V & N-body & none\\
 VI & N-body & none / backscaling \\
 \hline
\end{tabular}
\end{table}

The three cases (I)--(III) correspond exactly to the three different approaches discussed in the previous section that are designed to correctly include
linear radiation effects, and each of them is related to a respective gauge discussed in section \ref{sec:gauges}. The other three cases, (IV)--(VI) all
ignore the effect of early radiation , but they differ in the choice of gauge or initial conditions.
Hence the induced errors in the evolution will be different for these three cases.

\subsection{Radiation effect on matter power spectra}
\begin{figure*}
\includegraphics[width=\textwidth]{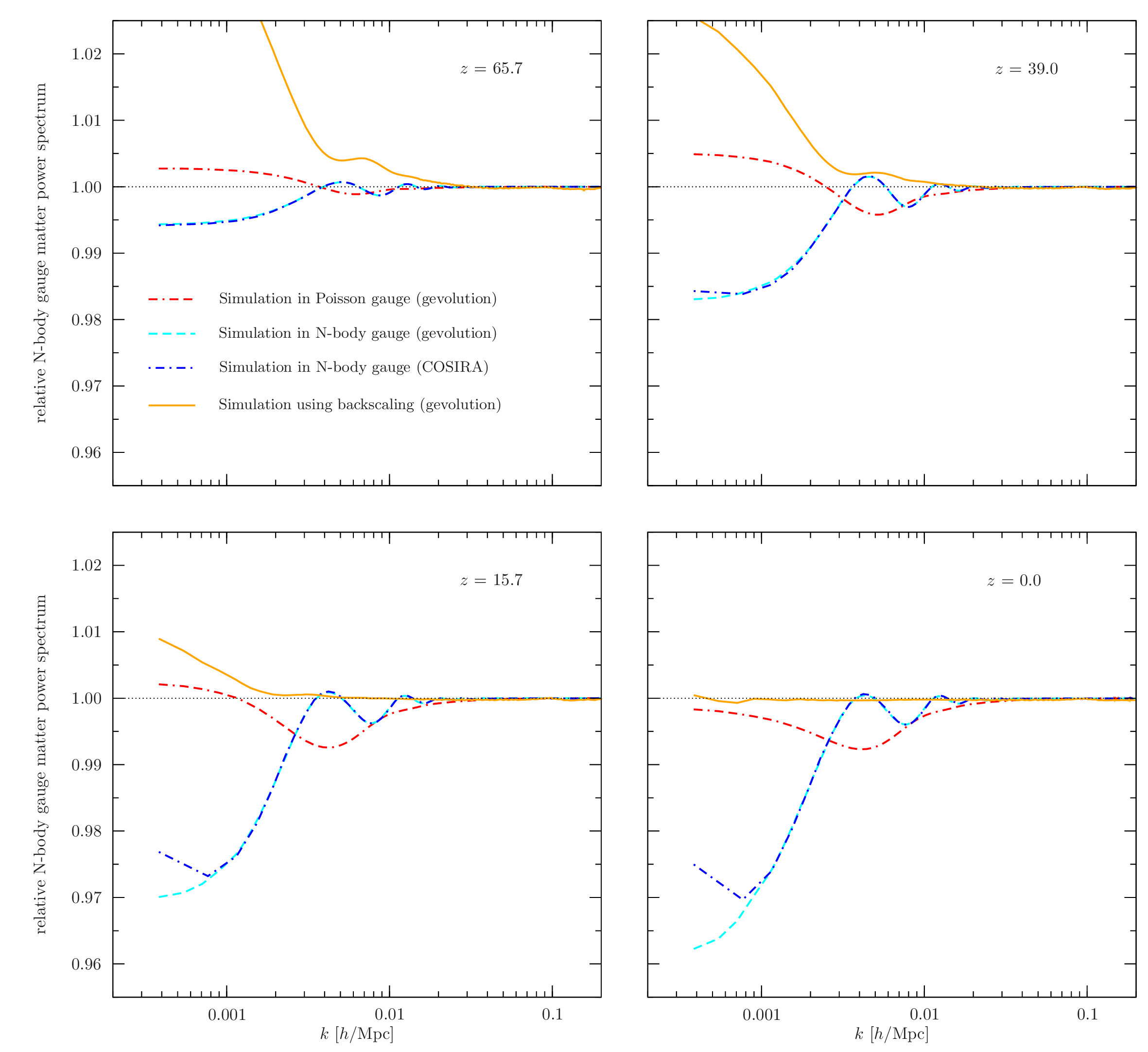}
\caption{
This figure shows the 
relative change in the Nb-gauge matter power spectrum at different redshifts 
for our simulations that \textit{did not} include radiation effects, compared to those that \textit{did} include them. For the red (dash-dash-dotted)
curve the simulations were carried out in Poisson gauge, case (II) and (IV) in the text, but we still show the ratio of the Nb-gauge power spectra. The light blue
(dashed) curve shows the analogous result if the simulations are 
performed directly in Nb gauge, case (I) and (V) .
A comparison with the dark blue (dot-dashed) curve shows that nearly identical results are obtained with \textsc{cosira}, and serves as a validation
of the two independent numerical implementations. The orange (solid) curve finally shows a simulation in Nb gauge where, instead of including radiation
in the dynamics, the initial data was manipulated in such a way that the error on the linear power spectrum is cancelled at redshift $z = 0$, a method called
``backscaling'' or case (VI) . 
}
\label{fig:Pratios}
\end{figure*}

We present our main results in figure~\ref{fig:Pratios} where we compare the scenarios without radiation,
cases (IV)--(VI), to the corresponding simulations that include radiation at four different redshifts. This illustrates the error introduced by
neglecting radiation perturbations in the various schemes. All realizations
use the same random numbers to set the perturbation amplitudes and phases,
such that the ratio of numerical spectra is not affected by cosmic variance.

In case (IV) the evolution is solved in Poisson gauge, but the contribution
of radiation to equations (\ref{eq:Pphi}) and (\ref{eq:chi}) is neglected. We compare this run (red, dash-dash-dotted curve)
against (II) which is a simulation where radiation is included.
The initial data is prepared using the linear transfer functions that
include radiation perturbations. The error is therefore zero at initial time, builds up to a certain amplitude at high redshift, and then essentially stops evolving
when radiation becomes more and more diluted.

Ignoring a physical effect
is not a gauge-invariant operation and it can therefore make a difference in which coordinate system one chooses to neglect radiation perturbations.
To illustrate this, we repeat the same exercise using Nb gauge for the evolution, where case (V) neglects the effects of radiation and case (I) includes them
(light blue, dashed curve).
We find that the error has a different shape and grows to about $4\%$ on the largest 
considered scales. For comparison,
we also plot the results that \citet{Brandbyge:2016raj} obtained with \textsc{cosira} (dark blue, dot-dashed curve). Apart from the first
data point which they already had noticed to be somewhat off we find a very good agreement between the two codes.

In our final example we discuss the common procedure to handle early radiation effects.
As noted above,
for case  (IV) and (V),  we  have
zero error on the power spectrum at the initial time, and consequently the improper handling of radiation causes the power spectrum to be offset from the true solution at later times. Given that most of the observations of large-scale structure are taken at low redshift, it actually seems more
natural to impose a vanishing error at late time, e.g. at redshift $z = 0$. Since early radiation affects only linear scales this can be achieved by the following
procedure.
First, compute the linear matter power spectrum at $z = 0$ using a Boltzmann code, fully taking into account radiation. Next, solve the linear mode equations
for matter perturbations for the given cosmological background, however, assuming that radiation has no perturbations at all. Finally, use  
the linear growing-mode solution for matter to scale the transfer functions from $z = 0$ back to the initialization redshift of the simulation.
This is the backscaling procedure that is commonly applied in the literature.

With this procedure the initial transfer functions will no longer agree with the relativistic ones, but they are designed in such a way that the
error introduced in the evolution by neglecting radiation is cancelled out at low redshift. A more detailed discussion of this method is presented in~\citet{Fidler:2017ebh}, where it is shown that this method works best using the present day Nb~gauge power spectrum. For this reason we compare the results obtained in scenario (VI) with the relativistic simulation (I) in the Nb~gauge. Our plot confirms that the error at the present time is vanishing, while at high redshifts a mismatch of several percent is found on large scales.

\subsection{Linear post-processing}
\begin{figure}
\includegraphics[width=\columnwidth]{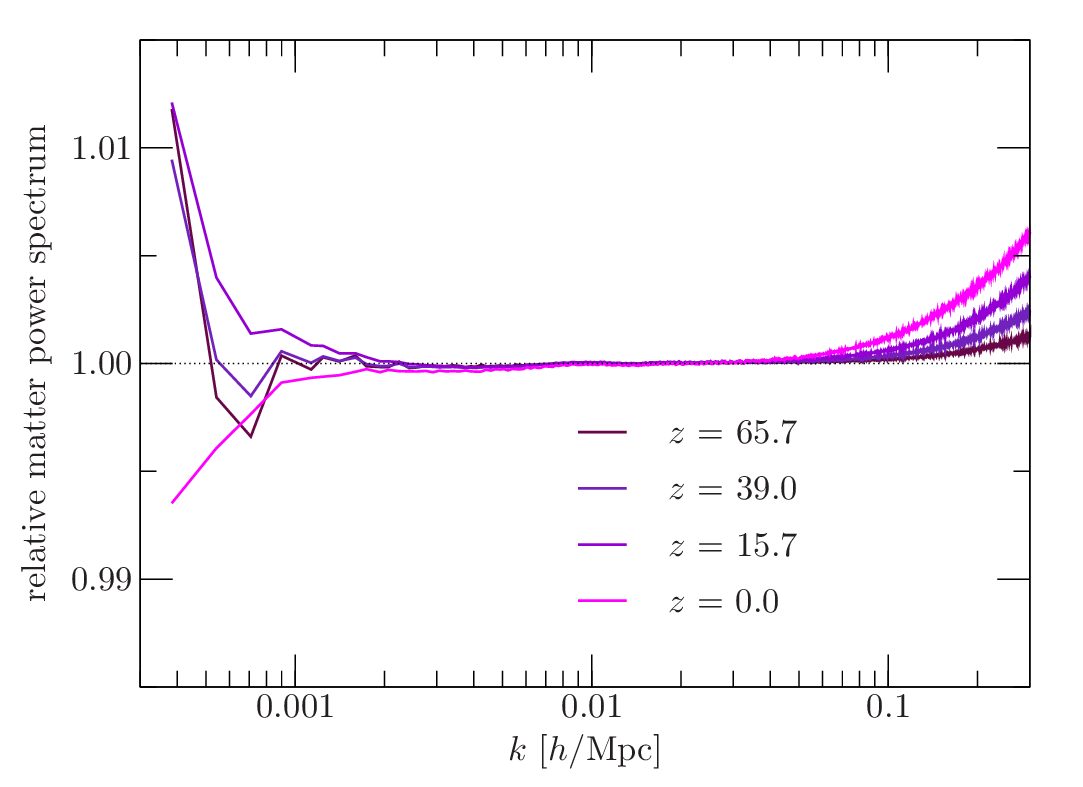}
\caption{Comparison of two simulations that both include radiation, case (I) and  (II) . The first simulation is performed directly in Nb gauge, whereas the second is performed in the Poisson gauge and subsequently converted to the Nb
gauge by acting a gauge transformation on the final particle positions. Since both simulations contain the same linear physics and the scales plotted are all in the linear regime, the ratio is unity up to errors introduced by the gauge transformation and the discretization.
}
\label{fig:PtoNb}
\end{figure}

A separate issue we want to discuss briefly is the numerical error that can be introduced by acting a gauge transformation on the final particle configuration.
By its nature, a linear gauge transformation is only applicable in the linear regime. However, in the
particular scenario studied in this article, the relativistic effects only appear on very large scales that remain linear to a high degree even down
to redshift $z = 0$. If we were to consider a situation where smaller scales would be affected, as could be the case in some inhomogeneous dark energy models
or models of modified gravity for example, the linear relation between Poisson and Nb gauge would be lost. In such a situation the Poisson gauge provides a framework where relativistic effects can be studied even at non-linear scales.
The Nb or Nm gauge framework are so far only defined to first order and it remains to be seen whether
they provide a useful concept in such an analysis.

Figure~\ref{fig:PtoNb} compares the scenarios (I) and (II) that both include radiation but are evolved in different gauges.
The ratio of power spectra is taken after the gauge transformation to Nb gauge, and we expect a ratio of unity if
physical results do not depend on the coordinate system used for the calculation.
Indeed, for the reasons noted above,
we find very good agreement between the simulations
in the two gauges. A small disagreement
of less than a percent is visible at the smallest scales, but this is probably due to discretization effects, as we see
the effect decreasing if we increase the resolution. One might wonder if the next-to-leading order weak-field effects, in particular frame dragging or the
anisotropic stress of dark matter, could also play a role on those scales. After all, these effects are taken into account in case (II) but are neglected
in case (I). However, these relativistic effects have a much smaller impact on the matter power spectrum, and are well below the permille level.

A somewhat larger disagreement between cases (I) and (II) appear at the largest scales and is of completely different origin. Here we are confronted with
the situation that the matter perturbations have very little power in Nb gauge, while the perturbations in Poisson gauge approach a nearly scale invariant
spectrum outside the horizon. This relatively large perturbation has to be taken off by the gauge transformation that takes the Poisson gauge to Nb gauge,
and achieving this to high precision can be difficult numerically. The numerical error is exacerbated by the fact that we are showing the relative power in
Nb gauge where the power itself is small. Taking all this into account we think that the agreement between cases (I) and (II) is very convincing.

As will become evident in the next subsection, the gauge transformation required in our scenario (III) that connects Nm gauge and Nb gauge is much less problematic, mainly due to the fact that the two gauges are much more closely related.

\subsection{Application of the Newtonian motion gauge}

\begin{figure}
\includegraphics[width=\columnwidth]{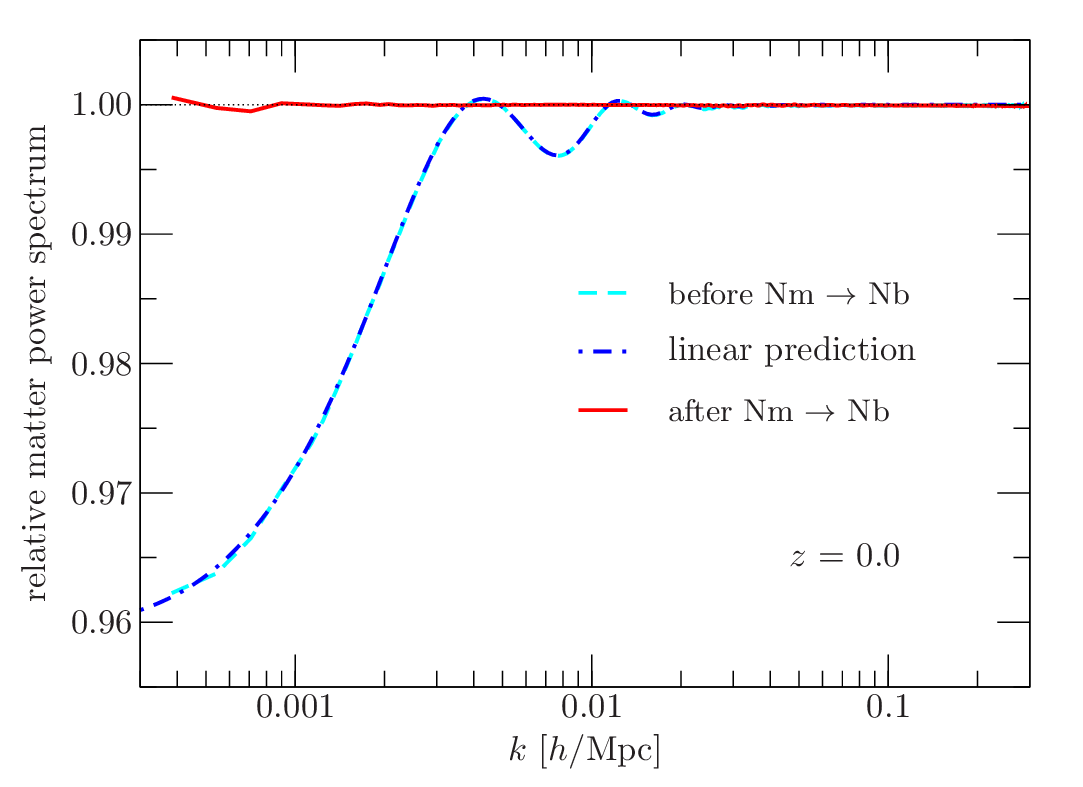}
\caption{The cyan (dashed) line shows the matter power spectrum at redshift $z=0$ of a Newtonian simulation that ignored radiation perturbations, relative
to a simulation that included those perturbations in Nb gauge. As shown this discrepancy is well described by a linear prediction  computed in \textsc{class} (blue, dot-dashed line). It can be accounted for by interpreting the result in terms of the Nm gauge.
To demonstrate this, the red (solid) line shows the matter power spectrum after the particles have been displaced according to the gauge transformation that
brings them from Nm gauge back to Nb gauge. The N-body simulations for this plot were
performed with \textit{gevolution}, and the linear transfer function
for the gauge generator was provided through a modified version of \textsc{class}.
}
\label{fig:Pnm}
\end{figure}

Scenario (III) that employs 
the Nm gauge does not fit into the above comparisons as it is 
non-nonsensical to neglect radiation in a Nm gauge. In figure~\ref{fig:Pnm} we compare the output of 
case~(III) before and after the gauge transformation to 
case~(I) in the Nb gauge. Before the gauge transformation, 
case~(III) is effectively identical to a Newtonian simulation and thus does not reproduce the results of the relativistic simulation (I). It does however match the comparison of the Newtonian power spectrum with the Nb gauge power spectrum at linear order performed in \textsc{class} to great accuracy. After displacing the particles according to the gauge transformation to the Nb gauge, we find a remarkable agreement with the results of simulation (I) with errors significantly below the percent-level on all examined scales. On small (very non-linear) scales, radiation effects are generally negligible, while on the larger scales the linear approximation of the Newtonian motion gauge metric potentials is a good approximation. 

Note that an alternative version of the Newtonian motion gauges defined in~\citet{Fidler:2017ebh} exists, that eliminates the error of the commonly employed backscaling method (VI) compared to the full Nb gauge simulation (I) by a similar gauge transformation.

\subsection{Anisotropic stress}\label{sec:aniso}

\begin{figure}
\includegraphics[width=\columnwidth]{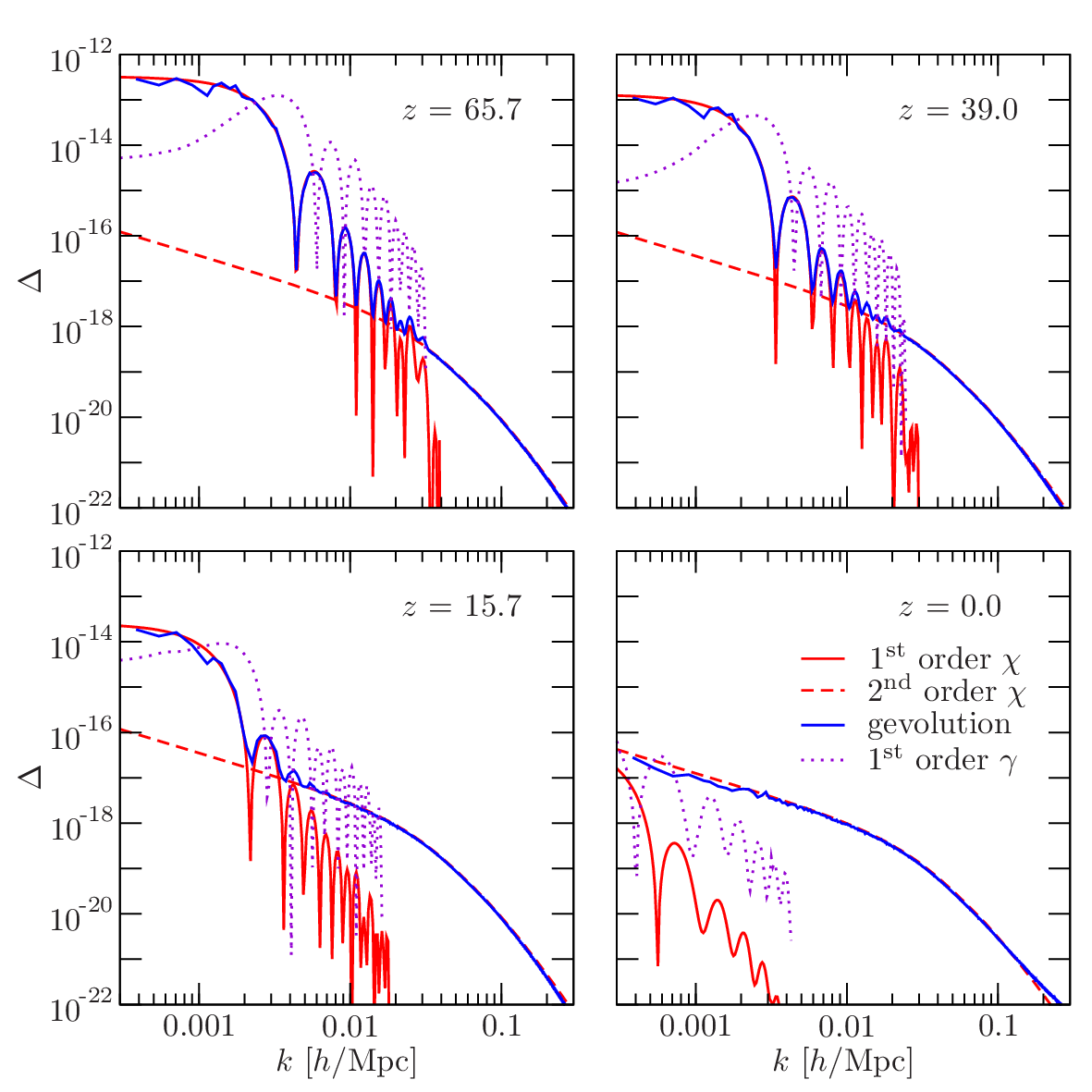}
\caption{Shown is the dimensionless power spectrum of $\chi$ at different redshifts.
The solid red line is the linear contribution
coming from radiation (photons and massless neutrinos). The dashed red line shows the second-order contribution from cold dark matter and geometry, as
computed in appendix \ref{app:chi}. The solid blue line is the one found in a numerical simulation with \textit{gevolution}. The dotted violet line shows
the dimensionless power spectrum of $\gamma$ for comparison.}
\label{fig:Pchi}
\end{figure}

A hallmark feature of relativistic physics is the presence of two distinct gravitational potentials. Free-streaming photons and neutrinos (which are taken massless in our study) 
generate  anisotropic stress even at
first order, sourcing $\chi$ according to equation (\ref{eq:chi}). This first-order contribution is gauge invariant and appears in Nb gauge as a contribution
to $\gamma$, see equation (\ref{eq:gamma}). The anisotropic stress of matter vanishes at first order, but it appears at higher order. While its
second-order contribution to $\chi$ can be computed perturbatively (see appendix \ref{app:chi}), \textit{gevolution} can construct it non-perturbatively
from the N-body ensemble. At this order, a new geometric contribution from the weak-field expansion has to be taken into account as well, see
equation (\ref{eq:chi2nd}). Since the Nb gauge framework is only developed up to leading order, we can do this computation so far only in Poisson gauge. 

Figure \ref{fig:Pchi} shows the dimensionless power spectrum of $\chi$,
defined in equation (\ref{eq:zetaspec}),
at four different redshifts.
We show separately the first-order contribution from radiation and the second-order contribution computed in appendix \ref{app:chi}, as well as
the total (first-order radiation plus second-order weak-field plus non-perturbative N-body contributions) found in the N-body simulation.
On the scales covered by the simulation we find excellent agreement with the perturbative results. Non-perturbative
effects would only show up on even smaller scales than what is resolved in these simulations. For comparison, we also show the power spectrum of $\gamma$, computed only
to first order since we have no second-order expression at hand. It turns out that $\gamma$ is typically about one order of magnitude larger than $\chi$ on
intermediate scales. Nevertheless, its (first-order) amplitude eventually drops below the second-order result for $\chi$, indicating that a first-order
calculation may no longer be meaningful beyond those scales. However for both, $\chi$ and $\gamma$, the overall amplitude decreases dramatically on small scales,
and their effect can eventually be neglected.

\section{Summary and conclusions}
\label{sec:conclusions}

We have presented a detailed study that investigates the effect of early radiation perturbations on the
matter power spectrum at various times. We have tested three methods that incorporate radiation, i.e., by the use of (1) the hybrid N-body code {\sc cosira},
(2) the relativistic N-body code {\it gevolution}, and (3) the framework of Newtonian motion gauges. The last method does not require a modification of the Newtonian N-body approach itself and includes, as a limit, the commonly applied method
to use modified initial data obtained through `backscaling'.

In more detail, the hybrid code {\sc cosira} modifies the N-body TreePM-code {\sc gadget-2} and couples it to the linear Einstein--Boltzmann code
{\sc class}. Radiation perturbations and GR effects are incorporated to the leading 
 order, and the N-body output is given in Nb gauge.
The second method is the relativistic N-body particle-mesh code {\it gevolution} which includes GR corrections up to second-order in the weak-field 
expansion in Poisson gauge. The linear effects of radiation are also  obtained by interfacing the code with \textsc{class}. Method~(3) is the Nm gauge framework that retroactively applies linear radiation and GR effects to unmodified Newtonian N-body simulations. Here the output of the simulation is to be interpreted within the Nm gauge, and the respective linear (metric) perturbations are determined with a modified version of {\sc class}.
This includes the method of backscaling where Newtonian simulations are initialized by using an appropriately rescaled present-day matter power spectrum.

To compare the various methods -- which make use of different gauges -- we transformed 
the simulation outputs to Nb gauge (if required).
For this we have applied an active linear gauge transformation on the final particle positions that displaces the
particles to the correct final positions in the Nb gauge. We have determined the required gauge generators by using a modified version of {\sc class}.

Our main results are summarized in figures~\ref{fig:Pratios}, \ref{fig:PtoNb} and~\ref{fig:Pnm}, where we take ratios of power spectra of the various methods that neglect/incorporate radiation. From figure~\ref{fig:Pratios} it is evident that the two relativistic codes {\sc cosira} and {\it gevolution} agree very well,
which gives us confidence that the two independent numerical implementations produce valid results. Only on very large scales there is a slight discrepancy between the codes which results from insufficient number of particles in the simulation carried out with {\sc cosira}.
Furthermore it is interesting to note that the results obtained in Poisson gauge (here by {\it gevolution}) are only very mildly affected by the presence of radiation perturbations, at most an effect of about one percent. In the Nb gauge, by contrast, the impact of radiation is up to four percent (for an initial redshift of $z_{\rm ini}=99$) on large scales. 
For a standard $\Lambda$CDM universe, the backscaling method works very accurately at the present time when interpreting the output in the Nb gauge, however, this simple interpretation is lost at the earlier times. A correct interpretation can still be achieved by employing the appropriate Nm gauge metric. In figure~\ref{fig:Pnm} we show our results obtained from the Nm gauge framework, where the impact of radiation is added to a Newtonian simulation as a post-processing. The results we obtain agree at better than percent-level accuracy on all studied scales.

In conclusion, we have established a comprehensive picture of the effects of early radiation in relativistic N-body simulations. These effects appear on very large scales
that can be treated linearly even at the present time. Using a relativistic 
framework
it is therefore  straightforward to take them into account, and
we have shown three different implementations that lead to similar results.
We have also verified that the backscaling approach accurately describes the matter power spectrum in the Nb gauge, with corrections from the full Nm metric relevant only on very large scales and
smaller than a few percent,
provided that the initialization redshift of the simulation is below $z \simeq 99$. Furthermore, we have verified that the backscaling method accurately simulates the evolution of the baryonic acoustic oscillation (BAO) feature. For this we have investigated the two-point matter correlation function for the various methods and found no significant discrepancy.

Perhaps it is relevant here to have a closer look at whether ignoring GR and radiation perturbations could potentially be an observable effect in upcoming surveys. Since the effect grows with increasing length scale it mainly affects surveys with large volumes, i.e.\ photometric surveys, rather than spectroscopic surveys which typically have smaller effective volumes but allow for very accurate reconstruction of the BAO feature. The largest such survey currently planned is the photometric redshift survey component of the \textsc{Euclid} satellite mission \citep[][]{Laureijs:2011gra}. As a very crude estimate of the precision with which the matter power spectrum can be probed at a given wave number $k$, 
we can use the approximate relation $\Delta P(k)/P(k) \sim \left[V_{\rm eff} k^3/(2 \uppi)^3 \right]^{-1/2}$ with an effective survey volume $V_{\rm eff}$ of approximately 1000\,Gpc$^3$ which yields $\Delta P(k)/P(k) \sim 0.015 \left(\frac{k}{0.01 \, h \, {\rm Mpc}^{-1}} \right)^{-3/2}$. At $k \sim  0.01 \, h \, {\rm Mpc}^{-1}$ we thus find $\Delta P(k)/P(k) \sim 0.015$ and at 0.001 we get $\Delta P(k)/P(k) \sim 0.5$. Clearly the effect of early radiation is at most marginally observable with \textsc{Euclid} data. However, future 21-cm surveys potentially have significantly larger effective volumes and in this case the effect is potentially both important and directly observable.

Finally, we note that 
the methods presented in this paper could be used
to study other types of (linear) relativistic effects that could occur, e.g., in models with inhomogeneous dark energy or other exotic sources of perturbations.

\paragraph*{Code availability:} A new release of \textit{gevolution} that includes the treatment of radiation effects is available on
a public {\sc git} repository.\footnote{\fontsize{7pt}{8pt}\selectfont \url{https://github.com/gevolution-code/gevolution-1.1.git}} The radiation module requires {\sc class}
to be linked as a library. We recommend to use the most recent public release.
The modified version of {\sc class} computing the Newtonian motion gauge potentials and gauge transformations is available upon request. 

\section*{Acknowledgements}

We thank Vincent Desjacques for useful discussions.
This work was supported by a grant from the Swiss National Supercomputing Centre (CSCS) under project ID s710.
CF is supported by the Wallonia-Brussels Federation grant ARC11/15-040 and the Belgian Federal Office for Science, Technical \& Cultural Affairs through the Interuniversity Attraction Pole P7/37.
CR\ is supported by the DFG through the SFB-Transregio TRR33 ``The Dark Universe''. 
JB, TT and SH acknowledge support from the Villum Foundation.

This is a pre-copyedited, author-produced version of an article accepted for publication in \textit{Mon.\ Not.\ Roy.\ Astron.\ Soc.} following peer review. The version of record, \textit{Mon.\ Not.\ Roy.\ Astron.\ Soc.\ 470 (2017) 303-313}, is available online at: \url{http://dx.doi.org/10.1093/mnras/stx1157}.

\bibliographystyle{mnras}
\bibliography{radiation}

\appendix

\section{Second-order contributions to  \texorpdfstring{$\chi$}{chi}}
\label{app:chi}

Deep inside the horizon and after the end of radiation domination the linear source terms for $\chi$ decay and the non-linear contributions
start to dominate. Here we present a perturbative calculation of these contributions that is accurate to second order and hence valid for mildly
non-linear scales \citep[see also][]{Ballesteros:2011cm}. To this end, we estimate the right-hand side of equation~(\ref{eq:chi2nd}) by inserting
the linear solutions that can be computed with a Boltzmann code. In particular, for non-relativistic matter, the stress is approximately given by
\begin{equation}
 \delta_{jk} T_i^k = \bar{\rho}\, \nabla_i v \nabla_j v \,,
\end{equation}
where $v$ is the linear velocity potential. Pressureless matter, even if it behaves like a perfect fluid, produces some anisotropic stress when you look
at it in a frame other than its rest frame. Defining a time-dependent density parameter
\begin{equation}
 \Omega = \frac{8 \uppi G a^2 \bar{\rho}}{3 \mathcal{H}^2} \,,
\end{equation}
and moving to Fourier space, equation~(\ref{eq:chi2nd}) becomes
\begin{multline}
 \chi_\mathbf{k} = \frac{1}{k^4 (2 \uppi)^{3/2}} \int \!\dd^3 \mathbf{q} \left[\phi_\mathbf{q} \phi_{\mathbf{k}-\mathbf{q}} + \frac{3}{2}\mathcal{H}^2 \Omega v_\mathbf{q} v_{\mathbf{k}-\mathbf{q}}\right] \\
 \times \left[3 \left(\mathbf{q}\!\cdot\!\mathbf{k}\right)^2 - 2 k^2 \left(\mathbf{q}\!\cdot\!\mathbf{k}\right) - k^2 q^2\right] \,.
\end{multline}
Note that we use the unitary Fourier convention.
The linear perturbation variables $\phi_\mathbf{k}$ and $v_\mathbf{k}$ are related to the initial amplitude of the gauge-invariant curvature perturbation $\zeta_\mathbf{k}^\mathrm{in}$ through linear transfer functions,
\begin{equation}
 \phi_\mathbf{k}(\tau) = \zeta_\mathbf{k}^\mathrm{in} T_k^\phi(\tau) \, , \qquad v_\mathbf{k}(\tau) = \zeta_\mathbf{k}^\mathrm{in} T_k^v(\tau) \,.
\end{equation}
Let us also define the dimensionless power spectrum $\Delta^{\!\zeta}(k)$ as
\begin{equation}
\label{eq:zetaspec}
 4 \uppi k^3 \left\langle \zeta_\mathbf{k} \zeta_\mathbf{k'} \right\rangle_{\rm c} = \left(2\uppi\right)^3 \delta_{\rm D}^{(3)}(\mathbf{k} + \mathbf{k'})\, \Delta^{\!\zeta}(k) \,,
\end{equation}
and similarly for other scalar quantities. With this definition and our Fourier convention the normalization of $\Delta(k)$ agrees with the one chosen by
\citet{Bernardeau:2001qr}. A straightforward calculation gives the following expression for the power spectrum of $\chi$,
\begin{multline}
\label{eq:chispec}
 \Delta^{\!\chi}(k) = \frac{1}{2\uppi k^5} \int\! \dd^3\mathbf{q} \left[T^\phi_q T^\phi_{|\mathbf{k} - \mathbf{q}|} + \frac{3}{2} \mathcal{H}^2 \Omega T^v_q T^v_{|\mathbf{k} - \mathbf{q}|}\right]^2 \\ \times \frac{\Delta^{\!\zeta,\mathrm{in}}(q)}{q^3}
 \frac{\Delta^{\!\zeta,\mathrm{in}}(|\mathbf{k} - \mathbf{q}|)}{|\mathbf{k} - \mathbf{q}|^3}
 \left[3 \left(\mathbf{q}\!\cdot\!\mathbf{k}\right)^2 - 2 k^2 \left(\mathbf{q}\!\cdot\!\mathbf{k}\right) - k^2 q^2\right]^2.
\end{multline}
At this point we would like to add a remark concerning the choice of metric parametrization. 
 It was noted by \citet{Adamek:2014xba} that a similar calculation
for a slightly different definition of $\chi$, related to the metric parametrization of \citet{Green:2011wc} (see also our footnote \ref{foot:metric}) that was followed also in \textit{gevolution}
prior to the latest version 1.1, gives an integral expression that is divergent in the infrared. This implies that the two-point correlation for $\chi$ (in that
parametrization) does not have a good fall-off behaviour at infinity. The divergence can be traced back to the appearance of a term
like $\delta_{ij} \phi \nabla^2 \phi$ in the equation that determines $\chi$. Such a term is absent in our corresponding equation~(\ref{eq:chi2nd}), and the integral
above therefore has no infrared problem. This is an additional motivation for choosing the exponential metric parametrization.

To clarify this point more explicitly, let us formally expand the potentials as
\begin{subequations}
\begin{align} 
  \phi &= \phi^{(1)} + \phi^{(2)} + \ldots \,, \\
  \psi &= \psi^{(1)} + \psi^{(2)} + \ldots \,, 
\end{align}
\end{subequations}
and let us do the same for the potentials $\Phi$ and $\Psi$ used by \citet{Adamek:2014xba}. Comparing the line elements, equation~(1) in their
work and our equation~(\ref{eq:Pmetric}), we establish
\begin{subequations}
\begin{align} 
  \Phi^{(1)} &= \phi^{(1)}\,, \\
  \Psi^{(1)} &= \psi^{(1)}\,, \\
  \Phi^{(2)} &= \phi^{(2)} - \left(\phi^{(1)}\right)^2\,, \\
  \Psi^{(2)} &= \psi^{(2)} + \left(\psi^{(1)}\right)^2\,.
\end{align}
\end{subequations}
Therefore $(\Phi - \Psi)$ and $(\phi - \psi)$ are the same at first order and correspond to the gauge-invariant quantity of equation (\ref{eq:chi}), whereas
they clearly differ at second order due to the reparametrization. Furthermore, the second-order contribution does not correspond to a gauge-invariant quantity
-- we simply compute it in a specific gauge. In this appendix we estimate $\chi^{(2)} = \phi^{(2)} - \psi^{(2)}$ which differs from $(\Phi^{(2)} - \Psi^{(2)})$
by a term quadratic in the first-order potentials. It is the latter which causes an infrared problem.

In order to evaluate the convolution integral of equation~(\ref{eq:chispec}) numerically, it is convenient to employ a variable transformation
$w = q/k$, $u = \sqrt{1 - 2 w \mu + w^2}$, where $\mu$ is the cosine of the angle between $\mathbf{k}$ and $\mathbf{q}$ \citep[e.g.][appendix C]{Lu:2008ju}. 
The azimuth angle can be integrated out directly, leaving us with
\begin{multline}
 \Delta^{\!\chi}(k) = \int\limits_0^\infty\! \frac{\dd w}{w^2}\!\! \int\limits_{|w-1|}^{w+1}\!\! \frac{\dd u}{u^2} \left[T^\phi_{kw} T^\phi_{ku} + \frac{3}{2} \mathcal{H}^2 \Omega T^v_{kw} T^v_{ku}\right]^2 \!\Delta^{\!\zeta,\mathrm{in}}(kw)\\ \times \Delta^{\!\zeta,\mathrm{in}}(ku) \left(\frac{1}{4} + \frac{1}{2} w^2 + \frac{1}{2} u^2 + \frac{3}{2} w^2 u^2 - \frac{3}{4} w^4 - \frac{3}{4} u^4\right)^2\,.
\end{multline}
As noted at the beginning of this section, this result is valid if the first-order contributions to $\chi$ are subdominant. Of course there exists a regime
where first and second-order contributions are of similar amplitude. In this case one can simply add the two contributions, noting that their cross correlation
has to vanish in perturbation theory if one assumes Gaussian initial conditions for $\zeta$.

\bsp	% typesetting comment
\label{lastpage}
\end{document}